\documentclass[showpacs,superscriptaddress,12pt]{revtex4-1} 
\usepackage{amsmath,amssymb,graphicx,bbm}
\usepackage{color}

\begin{document}

\title{Isolation of coherent and incoherent nonlinear spectroscopic signals by phase modulation }

\author{Khadga Jung Karki}\email{Khadga.Karki@chemphys.lu.se}
\affiliation{Chemical Physics, Lund University, Getingev\"agen 60, 222 41, Lund, Sweden}
\author{Loni Kringle}\affiliation{Department of Chemistry and Biochemistry, Univerisity of Oregon, Eugene, OR 97403-1253, US}
\author{Andrew H. Marcus}\affiliation{Department of Chemistry and Biochemistry, Univerisity of Oregon, Eugene, OR 97403-1253, US}
\author{T\~onu Pullerits}\affiliation{Chemical Physics, Lund University, Getingev\"agen 60, 222 41, Lund, Sweden}

\begin{abstract}
We investigate the effect of phase modulation of laser beams on the coherent and incoherent non-linear signals arising from the interaction of femtosecond pulses with matter. We observe that two collinear beams, whose phases are modulated by frequencies $\phi_1$ and $\phi_2$, produce two second harmonic signals from non-linear crystals whose intensities at the detector are modulated at the frequencies $\phi_2-\phi_1$ and $2(\phi_2-\phi_1)$. We also observe that an incoherent action signal, such as fluorescence and photocurrent, which arises from the absorption of two photons, is modulated at the same frequencies as in the case of second harmonic generation. We present a theoretical analysis to explain our observations. These results are important to understand how phase modulation techniques can be used to isolate different field-matter interaction pathways in a non-linear process. Because the method uses modulation of the signal intensity rather than wave-vector matching  to isolate different signals, it could be useful to perform multi-photon absorption studies on single molecules or nanoparticles.
\end{abstract}

\maketitle 

\section{Introduction}
Pulsed lasers, which are routinely used in spectroscopy, can have very high peak power. For example, the peak power of a 10 fs pulse with an energy of 1 nJ is on the order of 100 kW. When such pulses are focused to a spot size of 10 $\mu$m the peak intensity can reach $\sim$ 100 GW/cm$^2$. Interaction of pulses having such high intensity electric fields with a sample can generate a plethora of non-linear signals, which can be used to study various spectroscopic properties of the sample.\cite{MUKAMEL_BOOK,BOYD_BOOK} It is usually cumbersome, and also unnecessary, to investigate all of the non-linear signals in an experiment. Hence, different methods, such as phase matching, phase cycling and phase modulation, are often used to isolate the relevant signal contributions.\cite{HAMM_BOOK,WARREN_2003, MARCUS_2007} 

For an electric field described by a plane wave $E(t) = A(t) [\exp(i(\boldsymbol{k}x-\omega t))+\exp(-i(\boldsymbol{k}x-\omega t))]$, where $\boldsymbol{A}(t)$ is the temporal envelope of the field, the wave-vector $\boldsymbol{k}$ and the frequency $\omega$ quantify the momentum and energy of the photons, respectively. Conservation of energy and momentum in an $n$-wave mixing process\cite{BLOEMBERGEN_1980} leads to the generation of phase-matched non-linear signals with distinct wave-vectors $\boldsymbol{k}_j=\pm\boldsymbol{k}_1\pm\boldsymbol{k}_2\pm ... \pm \boldsymbol{k}_{n-1}$ and frequencies $\omega_j=\pm\omega_1\pm\omega_2\pm...\pm\omega_{n-1}$, where $\boldsymbol{k}_x$ and $\omega_x$ are the wave-vectors and the frequencies of the $x^{\textrm{th}}$ incident field. Phase-matching is used extensively to isolate specific field-matter interaction pathways in non-linear spectroscopy.~\cite{MUKAMEL_BOOK,BOYD_BOOK} For example, in a degenerate $4$-wave mixing (DFWM) experiment,~\cite{ZEWAIL_1996, JONAS_2003, TOKMAKOFF_2003, CHO_2008} when the frequencies of all of the beams are the same, one commonly changes the directions of the incoming beams (noncollinear geometry) to spatially isolate the signal of interest. 
Although DFWM experiments provide a wealth of information on different resonant and non-resonant properties of matter, they cannot be used to study isolated microscopic systems whose dimensions are smaller than the wavelength of the light; the uncertainty in the momentum of the photons emanating from a microscopic object prohibits spatial separation of the different non-linear signals from each other and also from the excitation beams. This problem could be circumvented by using phase cycling schemes in optical spectroscopy,\cite{WARREN_1981, WARREN_2003} in analogy to multi-dimensional NMR spectroscopy.\cite{ERNST_1976} In a typical phase cycling scheme, the relative phases between the different laser pulses are controlled by using pulse shapers and a set of measurements are done at different relative phases ( i.e., the relative phases in a single measurement are fixed). Addition of the spectra acquired at the well defined set of relative phases selects the desired non-linear signal.\cite{WARREN_2003,WARREN_2005,WARREN_2007} The relative phases between the pulses can also be changed continuously (dynamic phase cycling), as it is done in the technique of phase modulation,\cite{MARCUS_2006,MARCUS_2007} to separate the different signals arising from different field-matter interaction pathways.

Phase cycling and phase modulation techniques have mainly been used in optical two-dimensional action spectroscopy.\cite{WARREN_2003,WARREN_2005,WARREN_2007,MARCUS_2007,MARCUS_2011,MARCUS_2012} In these experiments, one detects the action of four field-matter interactions on the incoherent signal, such as fluorescence\cite{WARREN_2003,WARREN_2005,WARREN_2007,MARCUS_2007,MARCUS_2011,MARCUS_2012} or photocurrent.\cite{CUNDIFF_2013,KARKI_2014C} However, these techniques can be used more generally in isolating different coherent and incoherent signals arising from the non-linear interaction of the laser pulses with matter. Motivated by the generality of the technique and its anticipated use in the spectroscopic investigation of isolated micro and nano-sized systems and devices, we present in detail the concepts and principles on the use of phase modulation in non-linear spectroscopy. Similar principles can also be applied in  experiments that use phase cycling.

The polarization $\boldsymbol{P}$ induced in a medium by an electromagnetic field $\boldsymbol{E}$ can be written as \begin{equation}\label{EQ1}
\boldsymbol{P}(t) = \sum_{i=1}^{\infty} \chi_i \boldsymbol{E}^i(t),
\end{equation}
where $\chi_i$ is the $i^{\textrm{th}}$ order susceptibility of the medium. Although a propagating electric field has both spatial and temporal dependence, we ignore the spatial dependence in the following derivation in order to simplify the notation. This simplification is justified, as we use frequency rather than wavevector to separate the different non-linear signals in the phase modulation techniques. The total electric field due to a sequence of phase modulated pulses can be written as 
\begin{equation}\label{EQ2}
\boldsymbol{E}(t) = \sum_j \boldsymbol{A}(t) [\exp\{-i(\omega_j) t\}+\exp\{i (\omega_j) t\}],
\end{equation}
where $\omega_j=\omega+\phi_j$, and $\phi_j$ is the modulation frequency of the $j^{\textrm{th}}$ pulse. In Eq.\eqref{EQ2}, we have assumed that the amplitudes and the wavelengths of the different pulses (before the phase modulation) in the sequence are the same.\cite{Note1}   In a typical experiment, we use acousto-optic modulators (AOMs) to continuously sweep the relative phases of the optical fields. First-order diffraction from an AOM simply shifts the frequency of the incident laser beam, i.e. $\omega \mapsto \omega+\phi$. Here $\phi$ is the frequency of the acoustic-wave in the AOM. Note that one can equivalently choose another first-order diffraction geometry with $\omega-\phi$ for all the beams to achieve the same results. Substitution of $\boldsymbol{E}(t)$ in Eq.\eqref{EQ1} shows that the polarization induced by $n$ laser fields oscillates with frequencies $\pm\omega_1\pm\omega_2\pm...\pm\omega_n$. It is possible to continuously sweep the optical phases by other means, such as reflection from a vibrating surface or an electro-optic modulator. However, such methods produce undesired sidebands at the multiples of the modulation frequency; $\omega \mapsto \omega\pm n \phi$, where $ n\in \{0,\pm 1,\pm 2,..\}$. As we demonstrate below, this complicates the separation of different contributions to the non-linear signal.
 
 We provide examples that include sum-frequency generation, two-photon fluorescence, and two photon photocurrent  to illustrate how phase modulation can be used to separate the various signal contributions to nonlinear spectroscopic experiments.
 
\section{Phase modulated sum frequency generation}
\begin{figure}[htbp]
\includegraphics[width=13cm]{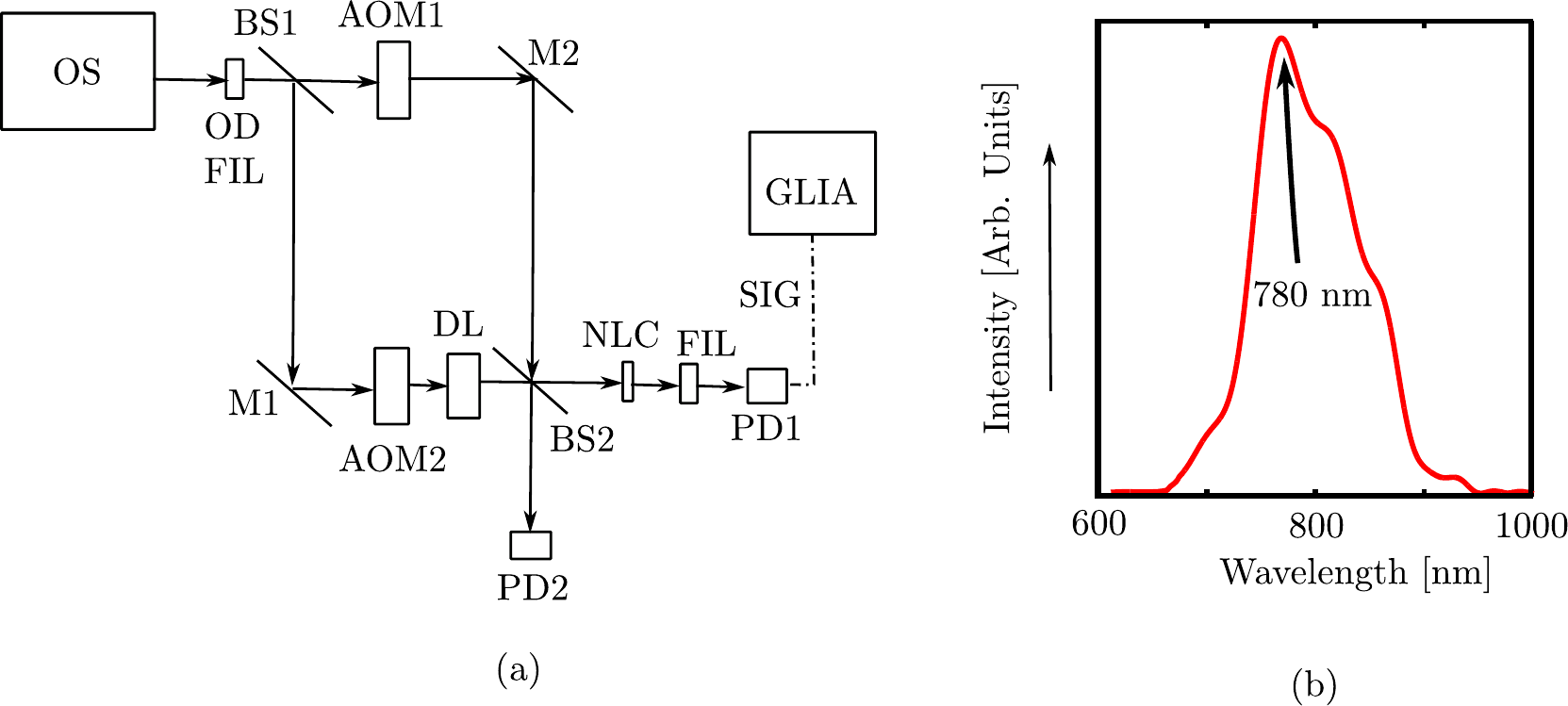}
\caption{(a) Schematics of the setup used to analyze the sum frequency generation using phase modulation. OS: oscillator, OD FIL: variable optical density filter, BS: beam splitter, AOM: acousto-optic modulator, DL: delay line, M: mirror, NLC: non-linear crystal, FIL: filter, PD: photodiode, SIG: signal and GLIA: generalized lock-in amplifier. The AOMs in the two arms of the Mach-Zehnder interferometer modulate the phases of the beams by $\phi_1$ and $\phi_2$, respectively. The two beams are recombined collinearly by BS2 and focused onto NLC. The second harmonic and sum frequency generated by NLC are spectrally filtered from the fundamental beams by FIL, and detected by PD. The signal from the PD is analyzed by GLIA. PD2 is used to record the intensity of the excitation beams. (b) Laser spectrum.}
\label{FIG1}
\end{figure}

In Fig.\ref{FIG1}(a), we show a schematic of the setup we used to investigate  sum frequency generation by phase modulation. A pulsed laser beam is split into two beams, each of which passes through one of the arms of a Mach-Zehnder interferometer. Each arm of the interferometer has an AOM that modulates the phase of the beam. The beams are then recombined collinearly, and focused onto a non-linear crystal (NLC). The temporal overlap between the pulses is optimized by using a delay line (DL). The second harmonic and sum-frequency signals from the NLC  are isolated from the fundamental beams by using a spectral filter (FIL), and detected by a photodiode (PD). The signal from PD is analyzed using a generalized lock-in amplifier (GLIA).\cite{KARKI_2013A,KARKI_2013C,KARKI_2014B}

The different second order non-linear processes induced in the NLC can be obtained from the expansion of the expression given in Eq.\eqref{EQ3}:
\begin{eqnarray}\label{EQ3} 
\boldsymbol{P}_2(t) &\propto& (\boldsymbol{E}_1(t)+\boldsymbol{E}_2(t))^2\propto (\exp\{-i(\omega+\phi_1)\}+\exp\{-i(\omega+\phi_2)\}+c.c.)^2\nonumber\\
&\propto& 2 +2[\exp\{-i(\phi_2-\phi_1)t\}+c.c]+2[\exp\{-i(2\omega+\phi_1+\phi_2)t\}+c.c.]\nonumber\\
&+& [\exp\{-i2(\omega+\phi_1)t+c.c.]+[\exp\{-i2(\omega+\phi_2)t\}+c.c.], 
\end{eqnarray}
where $c.c.$ stands for complex conjugate. The DC term and the terms with the frequency $\phi_2-\phi_1$ in Eq.\eqref{EQ3} are due to optical rectification. Note that the optical rectification in conventional second harmonic generation (SHG) only has the DC component. When phase modulation is used, the optical rectification also produces an AC field at frequency $\phi_2-\phi_1$. The terms with the frequency $2\omega+\phi_1+\phi_2$ are due to sum frequency generation (SFG), while the terms with the frequencies $2(\omega+\phi_1)$ and $2(\omega+\phi_2)$ are due to SHG from the individual beams. Only the contributions  due to SFG and SHG in Eq.\eqref{EQ3}, which propagate as electric field $\boldsymbol{E}_{\textrm{p}}$, are detected by the photodiode. The intensity of the signal, $I\propto\boldsymbol{E}_{\textrm{p}}^2 $, recorded by the photodiode is given by Eq.\eqref{EQ4}:
\begin{eqnarray}\label{EQ4}
I&\propto& [\exp\{-i4(\omega+\phi_1)t\}+c.c.]+[\exp\{-i4(\omega+\phi_2)t\}+c.c.]\nonumber\\
&+&3[\exp\{-i(4\omega+2\phi_1+2\phi_2)t\}+c.c.]+2[\exp\{-i(4\omega+3\phi_1+\phi_2)t\}+c.c.]\nonumber\\
&+& 2[\exp\{-i(4\omega+\phi_1+3\phi_2)t\}+c.c.]+[\exp\{-i2(\phi_2-\phi_1)t\}+c.c.]\nonumber\\
&+&4[\exp\{-i(\phi_2-\phi_1)t\}+c.c.].
\end{eqnarray}

Although Eq.\eqref{EQ4} has many oscillating terms, all of the terms that oscillate at high frequencies, i.e. the terms containing $\omega$, are filtered by the electronic circuit, so that only the terms with the frequencies  $\phi_{21}=\phi_2-\phi_1$ and $2\phi_{21}$  given in Eq.\eqref{EQ5} are detected by the GLIA:
\begin{equation}\label{EQ5}
I_{\textrm{det}}\propto [\exp(-i2\phi_{21}t)+c.c.]+4[\exp(-i\phi_{21} t)+c.c.].
\end{equation}
 It is perhaps counterintuitive that the detected non-linear signals oscillate at the relative modulation frequency, $\phi_{21}$, and twice the relative modulation frequency, $2\phi_{21}$. Although we have used a rather lengthy derivation to arrive at this relatively simple result, one can also use Feynman diagrams to reach the same conclusions. These diagrams also help us to rationalize why we observe the signals at the frequencies $\phi_{21}$ and $2\phi_{21}$. 

\begin{figure}[htbp]
\centering
\includegraphics[width=6cm]{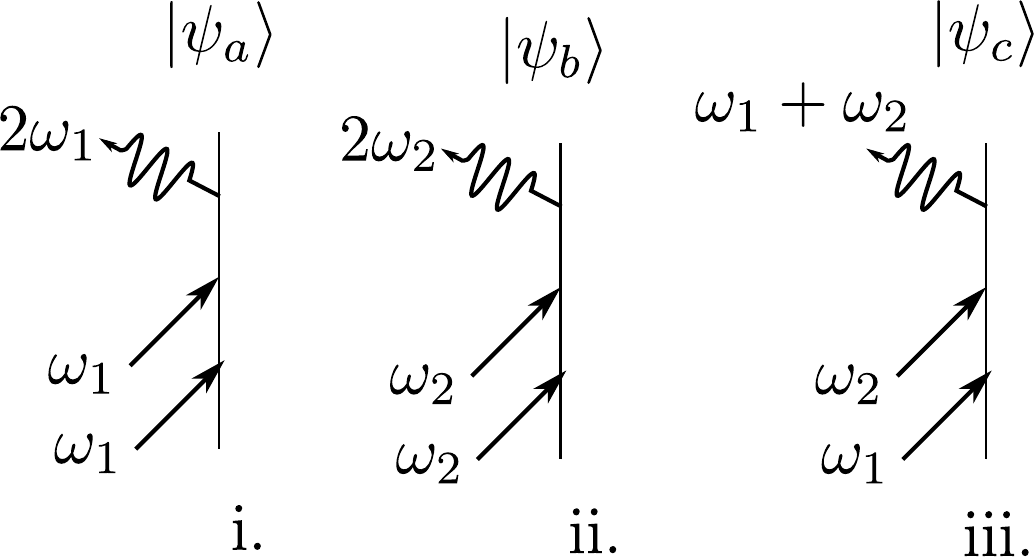}
\caption{Feynman diagrams of SHG (i, ii) and SFG (iii) processes. The field due to i. creates a wave-packet $|\psi_a\rangle$ and the field due to ii. creates a wave-packet $|\psi_b\rangle$ in the detector. The interference between the two wave-packets, $\langle \psi_b|\psi_a\rangle$, leads to the oscillation of the signal at 2$\phi_{21}$. Similarly, field due to iii. creates a wave-packet $|\psi_c\rangle$ in the detector. The interference of the wave-packets $\langle \psi_a|\psi_c\rangle$ and $\langle \psi_b|\psi_c\rangle$ generate the signal at $\phi_{21}$.}
\label{FIG_FND_2A}
\end{figure}

In Figs.\ref{FIG_FND_2A} (i., ii.), we show the Feynman diagrams of SHG processes that lead to the oscillation of the photocurrent from the PD at 2$\phi_{21}$. SHG processes described by pathways i. and ii. generate  wave-packets $|\psi_a\rangle\sim \exp\{-i2(\omega+\phi_1)t\}$  and $|\psi_b\rangle\sim \exp\{-i2(\omega+\phi_2)t\}$, respectively, in the PD. The interference between the two wave-packets $\langle \psi_b|\psi_a\rangle$ generates photocurrent with phase signature $\exp(i2\phi_{21}t)$. Note that the conjugate wave-packet interference term $\langle \psi_a|\psi_b\rangle$  produces a photo-current signal with phase signature  $\exp(-i2\phi_{21}t)$. These two signals combine to produce a real-valued photocurrent at $\exp(i2\phi_{21}t)+\exp(-i2\phi_{21}t)\propto \cos(2\phi_{21}t)$. Other interference terms  $\langle \psi_a|\psi_a\rangle$ and $\langle \psi_b|\psi_b\rangle$ produce DC signals, which can be removed using a high-pass filter preceding the GLIA. 

In Fig.\ref{FIG_FND_2A}(iii.), we show the Feynman diagram corresponding to the SFG process. Interference between the wave-packets generated by the SFG, $| \psi_c\rangle $, and SHG, $|\psi_a\rangle$, gives rise to the AC signal at $\phi_{21}$. Note that the interference of  $|\psi_c\rangle $ and $|\psi_b\rangle$  also contributes to the signal at the frequency $\phi_{21}$. Thus, the signals at $2\phi_{21}$ and $\phi_{21}$ allow us to separate the second-order non-linear processes of SHG and SFG, respectively. We further note that optimal wave-packet interference due to  the field-matter interactions of each beam requires that  the beams are collinear. 

We next demonstrate an experimental observation of the signals at frequencies $\phi_{21}$ and $2\phi_{21}$ that arise from SHG and SFG. We use a Synergy Ti:Sapphire oscillator from Femtolasers as the optical source in the setup [see Fig.\ref{FIG1}(a)]. The oscillator delivers 10 fs pulses at a repetition rate of 80 MHz. As example laser spectrum is shown in Fig.\ref{FIG1}(b). The spectrum is centered $\sim$780 nm and has a band-width of $\sim$135 nm. AOM1 is operated at $\phi_1=55$ MHz and AOM2 is operated at $\phi_2=55.17$ MHz. We used a 100 nm thick BBO crystal as the NLC. The SHG and SFG signals generated by the BBO crystal are filtered from the fundamental beams using a spectral filter (86-973, FL filter from Edmund optics). The filtered SHG and SFG signals at $\sim$370 nm are then detected using a GaP photodiode (PDA25K-EC, Thorlabs). We note that linear absorption of the fundamental beams by a low band-gap photodiode would also produce a signal at the relative modulation frequency $\phi_{21}$. Therefore, a spectral filter is necessary to separate the non-linear signal at the modulation frequency. In addition, as GaP does not absorb photons at the fundamental wavelength (700 -- 950 nm) the chance of detecting the linear signal becomes negligible. We also point out that the fundamental beams that leak through the filter are too weak to induce any two photon absorption in GaP. 

\begin{figure}[htbp]
\includegraphics[width=13cm]{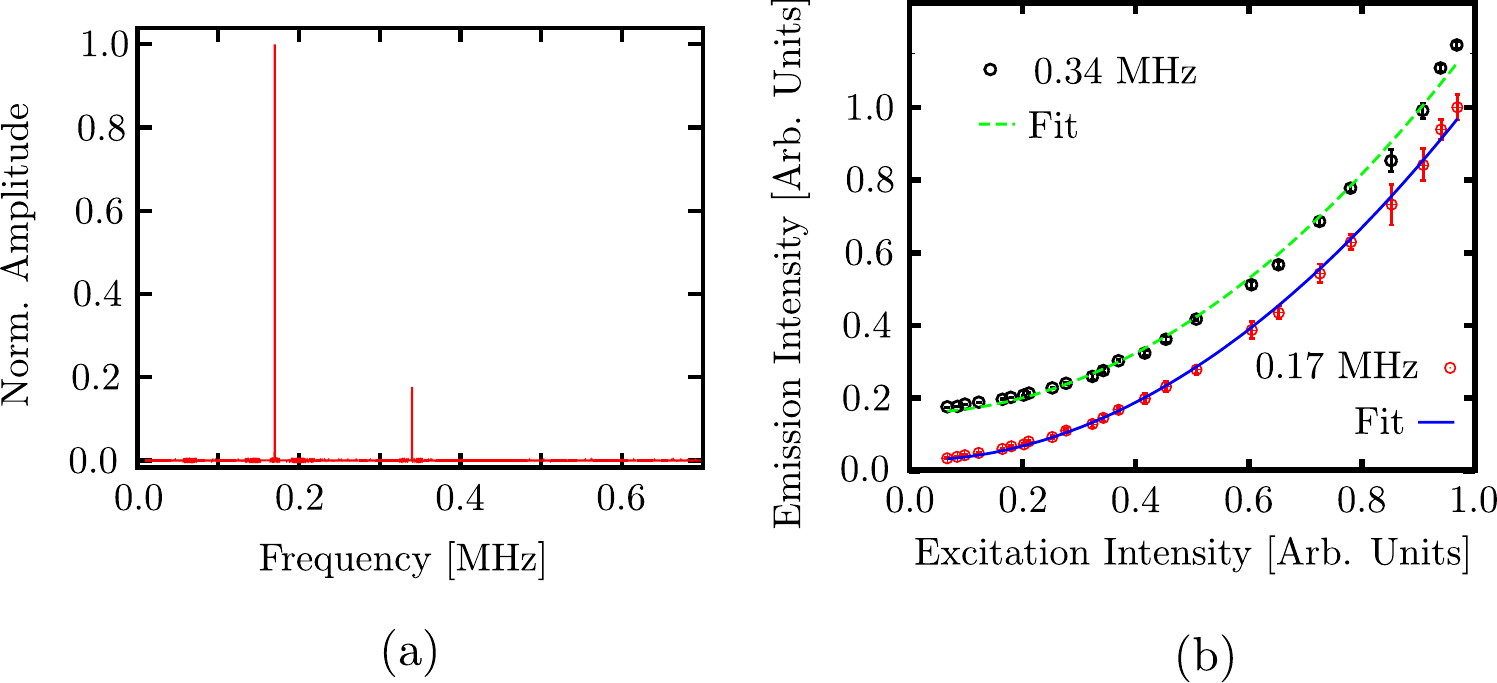}
\caption{(a): FFT of the SHG and SFG signals recorded by the photodiode, (b): excitation intensity dependence of the signal at $\phi_{21}=0.17$ MHz and $2\phi_{21}=0.34$ MHz. The points show the original data, and the lines are quadratic fits to the data points. }
\label{BBO_INT_DEP}
\end{figure}

In Fig.\ref{BBO_INT_DEP}(a), we show the fast Fourier transform (FFT) of the SHG and SFG signals produced by the BBO crystal, and recorded by the photodiode. The FFT shows two peaks, one at 0.17 MHz, which is the relative modulation frequency of the AOMs ($\phi_{21}$), and the other at 0.34 MHz (2$\phi_{21}$). The ratio of the amplitudes of the signals, $S_1/S_2$, is about 6:1, which is greater than 4:1, as predicted by Eq.\eqref{EQ5}. Two factors in the experimental conditions give rise to the deviation of the signal amplitudes from the ideal case given by Eq.\eqref{EQ5}. We have assumed that the amplitudes of the two beams, $\boldsymbol{A}_1$ and $\boldsymbol{A}_2$,  are equal [Eq.\eqref{EQ2}] in deriving Eq.\eqref{EQ5}. When $\boldsymbol{A}_1\neq \boldsymbol{A}_2$,  the ratio $S_1/S_2> 4$. This is also true when the two beams after the interferometer are not made to be perfectly collinear. In our experiments, $\boldsymbol{A}_1$ and $\boldsymbol{A}_2$ are not exactly equal, as the beamsplitters used do not have 50/50 splitting ratio over the entire spectrum of the laser. It is also challenging to maintain the identical transverse beam profiles after the AOMs. Hence, perfect spatial overlap of the beams has not been achieved in our experiments. 

The dependence of the signal intensities at 0.17 MHz and 0.34 MHz on the intensity of the laser beams are shown in Fig.\ref{BBO_INT_DEP}(b). Each data point was averaged over 20 measurements. The error bars in the data are the standard deviation of the different measurements.  We did not directly measure the intensity of the laser beams in the measurements. Instead, we measured the amplitude of the modulated signal at 0.17 MHz using a silicon photodiode (labeled PD2 in Fig.\ref{FIG1}). This signal scales linearly with the intensity of the laser beams. A variable OD filter (OD FIL in Fig.\ref{FIG1}) was used to change the intensity of the laser beams. The $x$-axis in Fig.\ref{BBO_INT_DEP}(b) is normalized by the maximum amplitude recorded by PD2, which corresponds to an average excitation power of about 10 mW (0.13 nJ per pulse) that was focused onto a spot size of $\sim$ 20 $\mu$m. The red points are the amplitudes of the non-linear signal recorded by PD1 at 0.17 MHz, and the blue line is the quadratic equation $f(x)=a+b(x-c)^2$ fit to the data points. Similarly, the  data  and the quadratic fit of the signal at 0.34 MHz are shown as black points and a green curve, respectively. The signal at 0.34 has been rescaled and shifted to fit in the plot. Both signals increase as a quadratic function of the laser beam intensity. These results show that the coherent second order non-linear signals induced by a pair of phase modulated collinear beams generate signals at $\phi_{21}$ and $2\phi_{21}$. 

\section{Phase modulated two-photon fluorescence}
The strength of a two-photon absorptive transition by a fluorophore is given by 
\begin{equation}\label{EQ6}
S_{2P}(t) \propto (\boldsymbol{E}_1(t)+\boldsymbol{E}_2(t))^4.
\end{equation}
Derivation of the full expression for $S_{2P}$ is similar to the derivation of $\boldsymbol{P}_2$. As in the case of SHG and SFG, we find that the intensity of the fluorescence induced by two-photon absorption is modulated at the frequencies $\phi_{21}$ and $2\phi_{21}$. In fact, the intensity of the detected two-photon fluorescence  is also given by Eq.\eqref{EQ5}. Again, we use Feynman diagrams to rationalize this apparent similarity.

\begin{figure}[htbp]
\centering
\includegraphics[width=9cm]{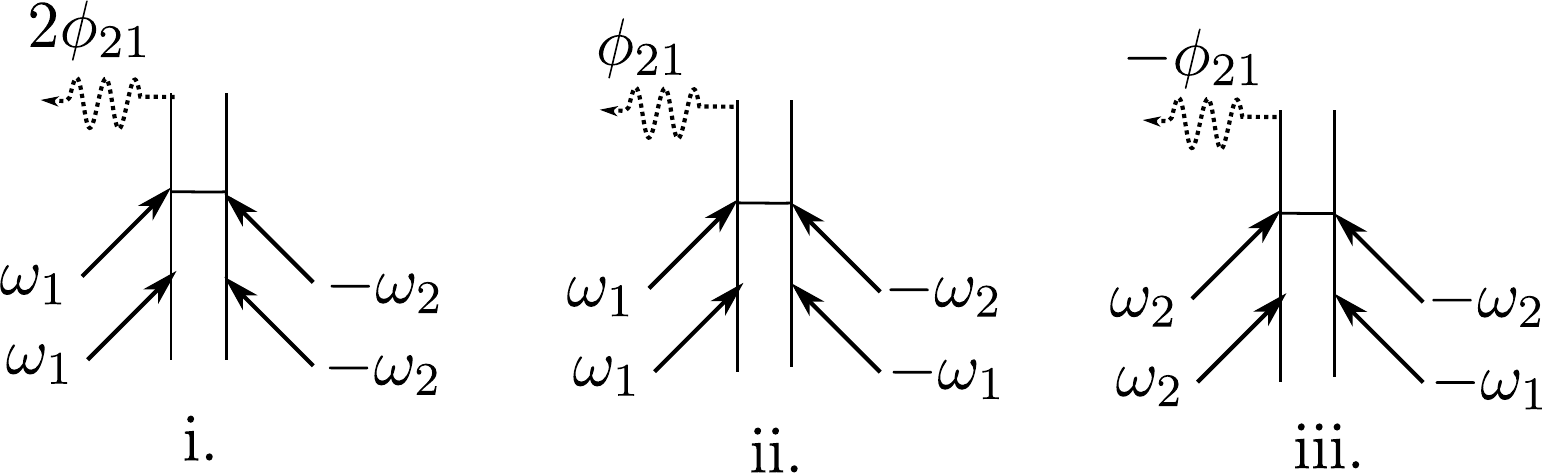}
\caption{Double sided Feynman diagrams of different pathways in two-photon fluorescence by phase modulation. Symmetric interaction of the fields from the two beams (i) gives rise to the fluorescence at 2$\phi_{21}$. Asymmetric interactions of the fields (three fields from one of the beams and one from the other) give rise to the fluorescence at $\phi_{21}$.}
\label{FIG_FND_2P2A}
\end{figure}
In Fig.\ref{FIG_FND_2P2A}, we show the Feynman diagrams of the different field interactions that give rise to two-photon fluorescence. The fluorescence intensity is modulated at the frequency 2$\phi_{21}$ when one photon from each of the two beams is absorbed by the fluorophores, as depicted in Fig.\ref{FIG_FND_2P2A}(i). When three fields from one of the beams, and the fourth field from the other beam participate in the absorption of two photons, the fluorescence is modulated at the frequency $\phi_{21}$. There are similarities between Fig.\ref{FIG_FND_2A}(i,ii) and Fig.\ref{FIG_FND_2P2A}(i). In the case depicted by Fig.\ref{FIG_FND_2A}(i,ii), the two pathways generate two coherent second harmonic signals, which interfere in the detector to generate the photocurrent at 2$\phi_{21}$. For the process shown in Fig.\ref{FIG_FND_2P2A}(i), the two signals interfere in the sample itself to generate intensity modulated fluorescence at the frequency $2\phi_{21}$. Similarly, the processes shown in Fig.\ref{FIG_FND_2A}(i,iii) and Fig.\ref{FIG_FND_2P2A}(ii), and Fig.\ref{FIG_FND_2A}(ii,iii) and Fig.\ref{FIG_FND_2P2A}(iii) produce analogous signals.

\begin{figure}[htbp]
\includegraphics[width=13cm]{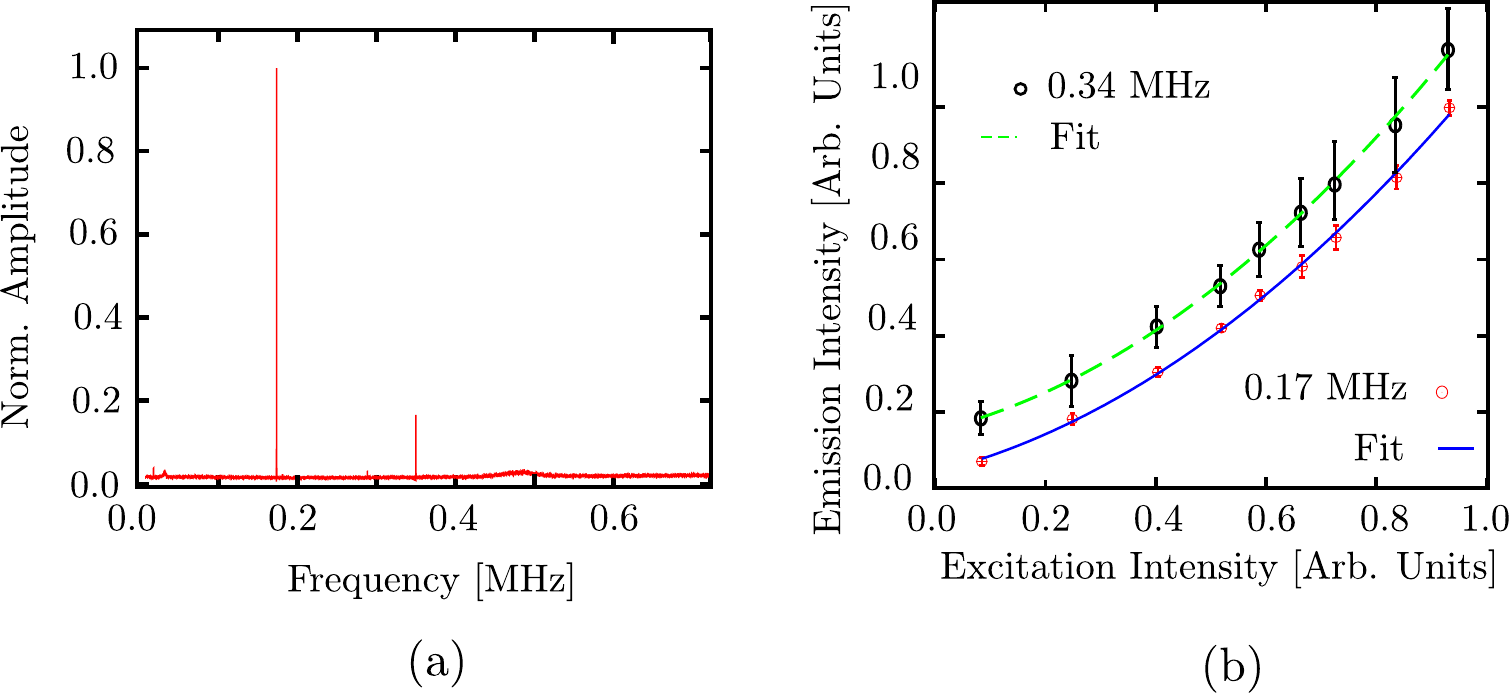}
\caption{(a): FFT of the two-photon fluorescence from the solution of Rhodamine 6G, recorded by the photodiode. (b): Excitation intensity dependence of the signal at $\phi_{21}=0.17$ MHz and $2\phi_{21}=0.34$ MHz. The points show the original data, and the lines are quadratic fits to the data. }
\label{RGH_INT_DEP}
\end{figure}

In Fig.\ref{RGH_INT_DEP}(a), we show the modulation frequencies of fluorescence signals resulting from a millimolar solution of Rhodamine 6G in water. The experimental setup is similar to that shown in Fig.\ref{FIG1}(a). For this experiment, the NLC is replaced by a cuvette with the solution of Rhodamine 6G sample. The two photon fluorescence is collected at 90$^{\circ}$ to the optical axis of excitation, and focused onto an avalanche photodiode (APD) (LCSA3000-01, Laser Components GmbH). A band-pass filter at 550 nm (84-772, Edmund Optics) was placed just before the APD to reject residual scattering of the fundamental beams. The FFT of the signal exhibits two sharp peaks at 0.17 and 0.34 MHZ, which correspond to the frequencies $\phi_{21}$ and $2\phi_{21}$, respectively. The ratio of the amplitudes at the two frequencies is about 6:1. As for the case of the coherent signal from the BBO crystal, the ratio deviates from 4:1. Here too, the deviation can be attributed to the lack of perfect spatial overlap of the beams due to the differences in the beam profiles.

In Fig.\ref{RGH_INT_DEP}(b), we show the amplitudes of the signals at 0.17 MHz (red points) and 0.34 MHz (black points) as a function of excitation intensity. The blue and the red lines are the quadratic functions fit to the data. We see that the fluorescence at both   the modulation frequencies increases quadratically with excitation intensity, which again demonstrates the nonlinear nature of these signals. Although these results are easily explained using Eq. \ref{EQ5}, they demonstrate an important application of phase modulation in non-linear spectroscopy. For example, only the signal at the frequency $2\phi_{21}$ has been used in multi-photon microscopy by phase modulation.\cite{WARREN_2009} Our results and analysis show that the non-linear signal at $\phi_{21}$ is more than four times stronger than the signal at $2\phi_{21}$. Therefore, in microscopy where the non-linear signal can be spectrally isolated from the linear signal, one can also monitor the signal at the modulation frequency $\phi_{21}$ to achieve a better signal-to-noise ratio.

\section{Phase modulated two-photon photocurrent}
\begin{figure}[htbp]
\includegraphics[width=13cm]{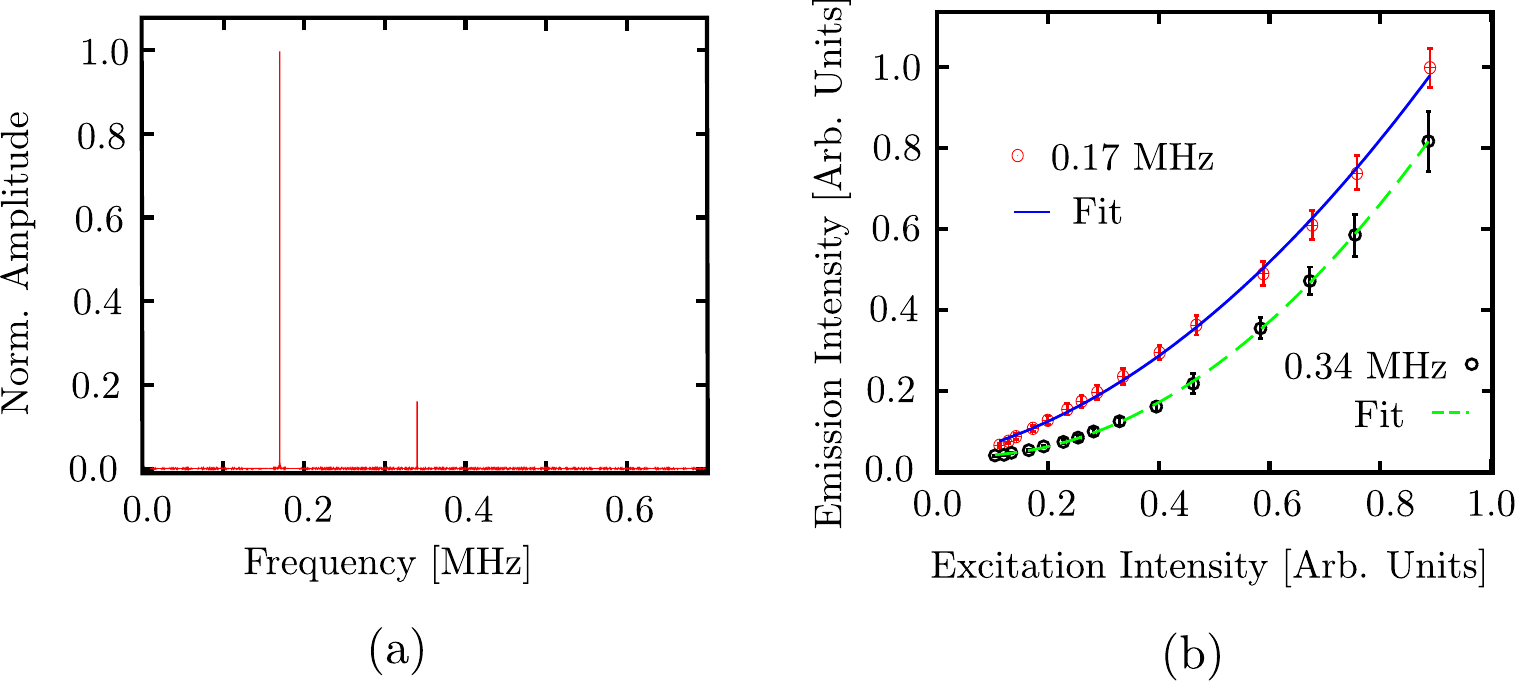}
\caption{(a): FFT of the two-photon photocurrent from GaP photodiode. (b): Excitation intensity dependence of the signal at $\phi_{21}=0.17$ MHz and $2\phi_{21}=0.34$ MHz. The points show the original data, and the lines are quadratic fits to the data. }
\label{GAP_INT_DEP}
\end{figure}

We next present an example of the use of phase modulation in investigating non-linear optical effects in devices, such as photodiodes.  As in the case of two-photon absorption induced fluorescence, when phase modulated beams interact with a two-photon photodiode, the resulting non-linear photocurrent is modulated at the frequencies $\phi_{21}$ and 2$\phi_{21}$. In Fig.\ref{GAP_INT_DEP}(a), we show the modulation frequencies of two-photon photocurrent from the GaP photodiode. These measurements were performed by replacing the NLC in the setup shown in Fig.\ref{FIG1}(a) by the GaP photodiode. The signals at 0.17 and 0.34 MHz are similar to the signals obtained from SHG and two-photon fluorescence. We observed that the signals at both of these modulation frequencies increased quadratically with the excitation intensity [see Fig.\ref{GAP_INT_DEP}(b)], which shows that both of the signals arose from the absorption of two photons by the photodiode.

\section{Discussion and Conclusions}
We have shown that modulation of the relative phase of two beams leads to the modulation of the intensity of the coherent and incoherent non-linear signals at multiples of the phase modulation frequency, $n\phi_{21}$. The amplitudes of the coherent signals due to the interaction of the system with three fields (second order non-linear process) are  modulated at the frequencies $\phi_{21}$ and $2\phi_{21}$. We have also shown that the amplitudes of the incoherent signals due to the interaction of the system with four fields are also modulated at $\phi_{21}$ and $2\phi_{21}$. Theoretical analysis shows that the amplitude of the signal with modulation frequency $\phi_{21}$ is four times larger than the amplitude of the signal with the frquency $2\phi_{21}$. In our experiments, we observe that the signal at $\phi_{21}$ is  roughly six times larger than the signal at $2\phi_{21}$. We attribute this discrepancy  to the imperfect overlap of the transverse beam profiles at the detector ( for the case of the coherent signals) and at the sample (for the case of the incoherent signals). Imperfect overlap occurs when the beams are not exactly collinear, and also when the modes of the beams do not exactly match with each other. 

Extension of the analysis of the field-matter interactions to higher orders reveals that the amplitudes of the coherent signals, which are due to $n-1$ field-matter interactions,  and the incoherent signals, which are due to $n$ field-matter interactions, are modulated at the frequencies given by $x\phi_{21}$, where $x=n/2, n/2-1, ..., 1$, and $n$ is an even integer. In the three examples presented in this work, we have not observed intensity modulation at frequencies higher than $2\phi_{21}$, and the signals scale quadratically with the excitation intensity. We have thus been able to observe only the interactions that involve at most four field-matter interactions. This limitation is due to the maximum intensity of the excitation beams available from our source laser oscillator. However,  laser beams from amplified systems should be able to drive even higher order non-linear processes. The phase modulation technique could be useful to the investigation of different interaction pathways in such high order processes. Moreover, as the technique can be used to investigate the different non-linear contributions to incoherent signals, the method could also be used to investigate the spectroscopic properties of isolated systems, such as individual molecules, nano-particles, and devices.

\end{document}